\documentclass[traditabstract]{aa}
\usepackage[dvips]{graphicx}
\usepackage{natbib}
\bibpunct{(}{)}{;}{a}{}{,}
\usepackage{txfonts}
\usepackage[mathscr]{eucal}
\begin{document}
\title{Evidence of grain growth in the disk of\\
       the bipolar proto-planetary nebula M 1--92}

\author{
	K. Murakawa\inst{1},
 	T. Ueta\inst{2},
	\and
        M. Meixner\inst{3}
	}


\institute{
	Max-Planck-Institut f\"{u}r Radioastronomie,
        Auf dem H\"{u}gel 69, D-53121 Bonn, Germany
	\and
        Department of Physics and Astronomy, University of Denver, Denver,
        CO 80208, USA
 	\and
        Space Telescope Science Institute, Baltimore, MD 21218, USA
	}


\abstract
{
We investigate the dust size and dust shell structure of the bipolar
proto-planetary nebula M 1--92 by means of radiative transfer modeling.
Our models consists of a disk and bipolar lobes that are surrounded by an AGB
shell, each component having different dust characteristics. The upper limit of
the grain size $a_\mathrm{max}$ in the lobes is estimated to be $0.5~\mu$m from
the polarization value in the bipolar lobe. The $a_\mathrm{max}$ value of the
disk is constrained with the disk mass (0.2~$M_{\sun}$), which was estimated
from a previous CO emission line observation. We find a good model with
$a_\mathrm{max}=1000.0~\mu$m, which provides an approximated disk mass of
0.15~$M_{\sun}$.  Even taking into account uncertainties such as the
gas-to-dust mass ratio, a significantly larger dust of
$a_\mathrm{max}>100.0~\mu$m, comparing to the dust in the lobe, is expected.
 We also estimated the disk inner radius, the disk outer radius, and the
envelope mass to be 30~$R_\star$(=9~AU), 4500~AU, and 4~$M_{\sun}$,
respectively, where $v_\mathrm{exp}$ is the expansion velocity.
If the dust existing in the lobes in large separations from the central star
undergoes little dust processing, the dust sizes preserves the ones in the dust
formation. Submicron-sized grains are found in many objects besides M 1--92,
suggesting that the size does not depend much on the object properties, such as
initial mass of the central star and chemical composition of the stellar system.
On the other hand, the grain sizes in the disk do. Evidence of large grains has
been reported in many bipolar PPNs, including M 1--92. This result suggests
that disks play an important role in grain growth.
}

\keywords{Stars: AGB and post-AGB -- circumstellar matter -- radiative transfer
-- polarization -- individual (M 1--92) }

\titlerunning{Grain growth in the disk of the bipolar PPN M 1--92}
\authorrunning{Murakawa et al.}
\maketitle

\section{Introduction}
Planetary Nebula (PN) morphology represents a history of physical processes
in the stellar$/$circumstellar environments. A large sample ($\sim$500) of
galactic PNs has shown that 20--25~\% of them have spherical structures, but
the others have asymmetric (elliptic, bipolar, and point-symmetric structures)
\citep[e.g.][]{za86,stanghellini93,cs95,manchado96,soker97,st98}. Several
theoretical studies have been carried out to explain the PN morphologies
\citep[see review by][]{bf02}. Widely accepted facts and physical processes are
cool spots on the stellar surface \citep{frank95,sc99}, the stellar rotation
\citep{bc93}, magnetic field \citep{pascoli92,mm98,mm99,matt00}, and binary
interaction \citep{lss79,morris81,morris87,bl90,sl94}. While details are still
unknown, these mechanisms one way or another cause the equatorially enhanced
mass loss in the AGB phase \citep[e.g.][]{meixner99,ueta00}. In the subsequent
post-AGB phase, the central star blows away the low density fast wind, which
inflates the lobe in the polar direction more than in the equatorial direction
\citep{kwok78,kwok82}. This interacting stellar wind (ISW) model explains
various morphologies of elliptic and bipolar PNs \citep{balick87}. In the above
scenario, the binary interaction is thought to be one of the most promising
mechanism to shape PNs with a narrow waist between the bipolar lobes because
this can amplify the mass loss in the equatorial plane and make a disk-like
structure \citep[e.g.][]{sr00}. In addition, the binary interaction is also
thought to form asymmetric structures such as jets, ansae \citep[e.g.][]
{soker90,soker92}, and point-symmetric shapes such as a spiral and quadruple by
precessing motion \citep{manchado96,mh06}.

It is difficult to directly detect the effect of binary interaction on PN
shaping using the current observing techniques. However, it would be possible
to infer the presence of the effects due to the aforementioned scenario if the
inner part of the circumstellar environment is investigated closely enough.
Near-infrared imaging polarimetry is a powerful method to probe dust shells and
to provide important information that can be derived in more details than
non-polarimetric imaging. Previous observations and dust scattering modeling
have revealed disk and bipolar lobe structures of evolved stars
\citep[e.g.][hereafter UMM07]{ss95,su03,gledhill05,ueta05,murakawa05,murakawa07,ueta07}.
However, most previous experiments treated only a single grain model in the
entire dust shell or SEDs were not considered simultaneously. These shortcomings
have prohibited us from investigating into the grain and disk properties in
detail. To resolve the situation, we considered both polarimetric images and
SEDs in our recent radiative transfer modeling of the bipolar proto-PN (PPN)
\object{Frosty Leo}, assuming two different dust models, one in the equatorial
region (disk) and the other in the bipolar lobes \citep{murakawa08a}. From this
modeling, we were able to derive evidence for grain growth in the disk region.
In this paper, we use this technique for the bipolar PPN: \object{M 1--92}
(=IRAS 19343+2926), whose characteristics were discussed previously by us
(UMM07). In Sect.~\ref{rtmodel}, details of our radiative transfer modeling are
described. We will discuss the implication of our model result in
Sect.~\ref{discuss}.

\section{Radiative transfer calculations}\label{rtmodel}
The primary purpose of our modeling is to explore the parameter space of the
disk geometry and the grain sizes in the circumstellar dust shells of our
target. In order to do this, we applied model geometries with a disk and bipolar
lobes and different dust models for them. We used our own three-dimensional
Monte Carlo code \textsf{STSH} \citep{murakawa08b}, which solves radiative
transfer problem of scattering and absorption by dust. This code can handle
multiple dust models in an arbitrary model geometry and computes SEDs, dust
temperatures, and the Stokes $IQUV$ parameters. We constrained the model
parameters by comparing with the observed SEDs and polarimetric images from the
HST NICMOS 2 data at a wavelength of 2~$\mu$m. We followed a similar modeling
procedure as in the past \citep{murakawa08a,murakawa08b}. We first
tried several parameter sets to find an approximated solution and parameter
ranges and determined some parameters, which can be estimated easily.
The grain sizes in the bipolar lobe were determined from the polarization
values in this region, which do not depend on the other parameters such as the
model geometry and the masses of the disk and envelope. Then, we constructed
a large number of models by SED fit and made model images of a few possible
parameter sets. We find some good models with different dust sizes
$a_\mathrm{max}$ in the disk, which reproduces the characteristics of the above
observations well. To constrain the $a_\mathrm{max}$ value, we use the disk
mass which was estimated from a previous CO emission line observation
\citep{bujarrabal98b,alcolea07}. We will describe this grain size issue in
detail below.

\subsection{Model assumption and numerical simulation}\label{i19343}
M 1--92 is an oxygen-rich PPN known as Minkowski's footprint
\citep{minkowski46}. An optical spectrum suggests a 6500~K black body component
and an 18\,000~K component in shorter wavelengths. The latter is probably due to
flux from the companion star \citep{arrieta05}. Because the 6500~K component
dominates in the total flux, we assumed a blackbody spectrum with this
temperature as the illumination source. The distance $D$ was determined to be
between 2.5~kpc \citep{ck77} and 3.5~kpc \citep{eh89}. We adopted 3.0~kpc, the
average of the two. For the luminosity, although \cite{ck77} obtained a
1600~[D kpc]$^2~L_{\sun}$, which corresponds to $\sim$20\,000~$L_{\sun}$ at
3.5~kpc, we found that 7000~$L_{\sun}$ fits the absolute flux in the SED better
and adopted this value in our modeling.

\begin{table}
  \begin{center}
  \caption[]{Model parameters of our radiative transfer calculations.}
  \label{i19343_parm}
  \begin{tabular}{lll}
  \hline
  \hline
  parameters        & adopted values           & comments$^1$ \\
  \hline
  \multicolumn{3}{c}{central star} \\
  $T_\star$         & 6500~K                   & $^2$         \\
  $L_\star$         & 7000~$L_{\sun}$          & adopted$^3$  \\
  $d$               & 3.0~kpc                  & $^4$         \\
  $R_\star$         & $1.3\times10^{12}$~cm    & calculated   \\
  \hline
  \multicolumn{3}{c}{disk} \\
  $R_\mathrm{in}$   & 30~$R_\star$(= 9~AU)     & 30 -- 50     \\
  $R_\mathrm{disk}$ & 4500~AU                  & 3000 -- 4500 \\
  $H$               & 0.3                      & 0.3 -- 0.4   \\
  $\tau_\mathrm{2.0}$ & 30                     & adopted$^3$  \\
  $M_\mathrm{disk}$ & 0.15~$M_{\sun}$          & $^5$         \\
  $a_\mathrm{max}$  & 1000.0~$\mu$m            & $\ga10.0$    \\
  $f_\mathrm{m}$    & 0.2                      & adopted$^3$  \\
  \hline
  \multicolumn{3}{c}{superwind and AGB shell} \\
  $R_\mathrm{sw}$   & 7\farcs5                 & adopted$^6$  \\
  $\alpha$          & 1.2                      & adopted$^6$  \\
  $\beta$           & 1.7                      & adopted$^6$  \\
  $\gamma$          & 0.7                      & adopted$^6$  \\
  $\epsilon_\mathrm{in}$  & 0.01               & adopted$^6$  \\
  $\epsilon_\mathrm{rim}$ & 0.5                & adopted$^6$  \\
  $R_\mathrm{out}$  & 30\arcsec                & 30 -- 60     \\
  $a_\mathrm{max}$  & 0.5~$\mu$m               & adopted$^7$  \\
  $f_\mathrm{m}$    & 0.2                      & adopted$^3$  \\
  $M_\mathrm{env}$  & 4~$M_{\sun}$             & 4 -- 6       \\
  $\dot{M}$         & \multicolumn{2}{l}{7.5$\times$$10^{-6}[v_\mathrm{exp}$ kms$^{-1}]~M_{\sun}$yr$^{-\mathrm{1, 8}}$} \\
  \hline
  \end{tabular}
  \end{center}
  $^1$ Ranges give the uncertainty of the corresponding model parameters,
  $^2$ \cite{arrieta05}, $^3$ Based on comparison of the SED,
  $^4$ \cite{ck77,eh89}, $^5$ with assumption of a $a_\mathrm{max}=1000.0~\mu$m
  grain model, $^6$ Based on comparison of the intensity image, $^7$ Based on
  comparison of the polarization image, $^8$ calculated.
\end{table}

\begin{figure*}
  \centering
   \resizebox{\hsize}{!}{\includegraphics{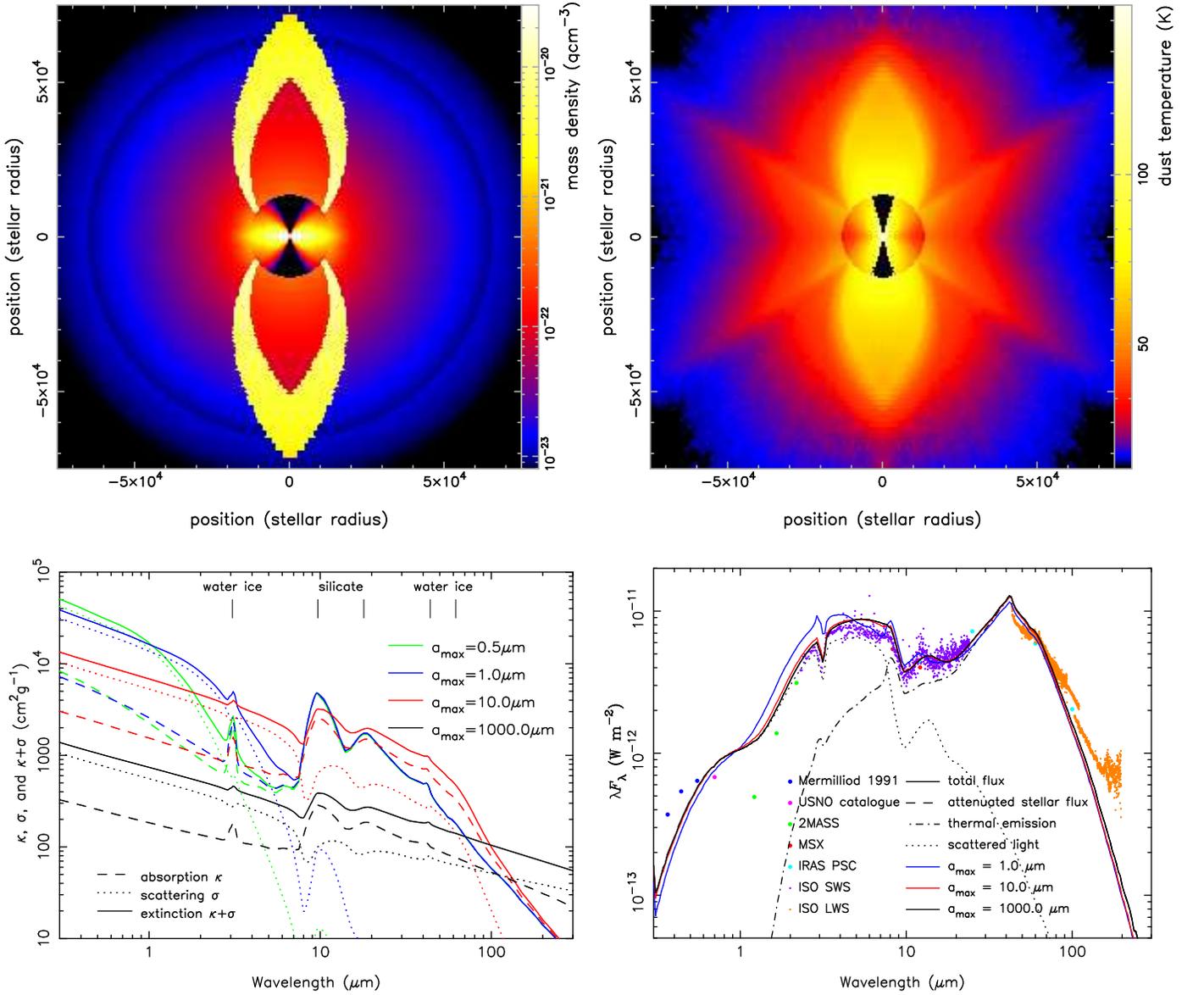}}
  \caption{Mass density distribution of the selected model (top left panel),
           dust temperature distribution (top right panel), dust opacities as
           function of wavelength (bottom left panel), and comparisons of model
           SEDs with observations (bottom right panel). For the selected model,
           the attenuated stellar flux, thermal emission, and scattered light
           from the dust are also plotted with the dashed-curve,
           the dashed-dotted curve, and the dotted curve, respectively.
           The observed data are from \cite{mermilliod94} and USNO catalog for
           optical photometry, point source catalogs of 2MASS, MSX, and IRAS
           for infrared photometry, and ISO SWS and LWS for mid- and
           far-infrared spectra, respectively. In the opacity plot, the mean
           opacities of absorption $\kappa$, scattering $\sigma$, and extinction
           $\kappa+\sigma$ are indicated with dashed, dotted, and solid lines,
           respectively.  Models with four different $a_\mathrm{max}$ values of
           0.5~$\mu$m, 1.0~$\mu$m, 10.0~$\mu$m, and 1000.0~$\mu$m are compared.
           The $a_\mathrm{max}=0.5~\mu$m model, which is used in the superwind
           and AGB shells, is also presented for comparison. In the SED plots,
           several curves with different colors denote model results with
           different grain sizes ($a_\mathrm{max}$) in the disk.
         }
  \label{rts_i19343}
\end{figure*}

\begin{figure*}
  \centering
  \resizebox{\hsize}{!}{\includegraphics{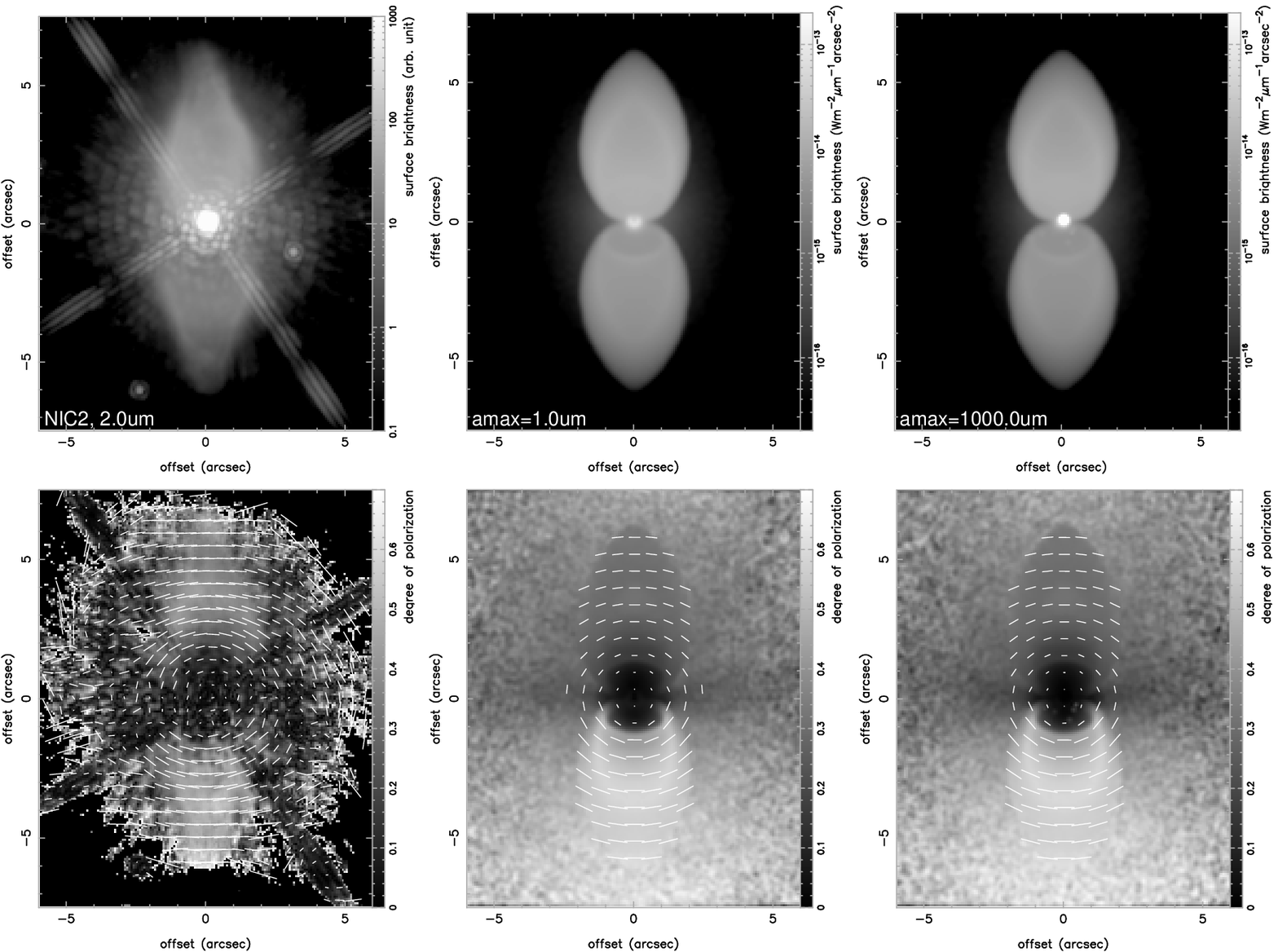}}
  \caption{Model results of the intensity (upper panels) and polarization
           (bottom panels) images of M1--92 comparing with the HST$/$NICMOS 2
           data from UMM07 (left column). Models with grain sizes in the disk
           of $a_\mathrm{max}=1.0~\mu$m (middle column) and
           $a_\mathrm{max}=1000.0~\mu$m (right column) are presented to show the
           effects of grain sizes. In the polarization images, the polarization
           vector lines are also plotted.
         }
  \label{img_i19343}
\end{figure*}

The nebula has a waterlily petal-like shape with an $11\arcsec\times6\arcsec$
extension \citep[see also][]{tg96,bujarrabal98a}, which extends towards the
northwest and the southeast directions. The northwestern lobe is brighter than
the southeastern one. Blue- and red-shifted components are detected in the
northwestern and southeastern lobes, respectively, in the CO emission line data
and the northwestern lobe is tilted towards us by $35\degr$
\citep{bujarrabal98b}. In our modeling, we assume that the model geometry
consisting of three components: (1) a disk in the innermost part,
(2) a superwind shell in the outside of the disk, and (3) an AGB dust shell in
the outermost part. The disk form chosen is one studied by \cite{toomre82} and
\cite{fischer96}. For the superwind shell, the NICMOS~2 intensity image shows
that the surface brightness at the outline of the bipolar lobes is slightly
enhanced compared to the projected inner part \citep{ueta07}, suggesting
a hollow structure, which is probably formed by the interaction with the ISW.
The AGB shell is assumed to have a one-dimensional, spherically symmetric
structure with a single power-law radial density gradient. The mass density
distribution $\rho$ consists of the individual components of the disk
$\rho_\mathrm{disk}$, the superwind shell $\rho_\mathrm{sw}$, and the AGB shell
$\rho_\mathrm{AGB}$ and is given by
\begin{eqnarray}
  \rho&=&\rho_\mathrm{disk}+\rho_\mathrm{sw}+\rho_\mathrm{AGB},\nonumber\\
  \rho_\mathrm{disk}\left(r,z\right)&=&
    \rho_\mathrm{d}\left(r/R_\mathrm{in}\right)^{-2}
    \exp\left[-\left|\frac{z}{Hr}\right|\right]
    \mbox{ for $R_\mathrm{in}\le R\le R_\mathrm{disk}$},\nonumber\\
  \rho_\mathrm{sw}\left(R,\theta\right)&=&
    \rho_\mathrm{e}\left(R/R_\mathrm{sw}\right)^{-\alpha}
    \left\{
      \begin{array}{@{\,}ll}
        \epsilon_\mathrm{in} & \mbox{ for $R_\mathrm{disk}\le R\le\gamma R_\mathrm{lobe}$},\nonumber\\
        \epsilon_\mathrm{rim} & \mbox{ for $\gamma R_\mathrm{lobe}\le R\le R_\mathrm{lobe}$},\nonumber\\
      \end{array}
    \right.\nonumber\\
  \rho_\mathrm{AGB}\left(R\right)&=&
    \rho_\mathrm{e}\left(R/R_\mathrm{sw}\right)^{-2}
    \mbox{ for $R_\mathrm{sw}\le R\le R_\mathrm{out}$},\nonumber
\end{eqnarray}
where the coordinates of $\left(R,\theta\right)$ and
$\left(r,z\right)=\left(R\sin\theta,R\cos\theta\right)$ are the two-dimensional
spherical coordinate and cylindrical coordinate, respectively. We assume that
the disk exists in a region between the inner radius $R_\mathrm{in}$ and the
outer radius $R_\mathrm{disk}$. The value $H$ is the ratio of the disk height
to the disk radius. The $\rho_\mathrm{d}$ is the density coefficient and is
derived from the optical depth in the equatorial plane instead of the disk mass
$M_\mathrm{disk}$ as done before \citep{meixner02,murakawa08a}. We find that an
optical depth of 30 at a wavelength of 2.0~$\mu$m fits well in the optical to
NIR fluxes. We apply this value for further investigation of the grain size
effect. The shape of the superwind is determined by
$R_\mathrm{lobe}=R_\mathrm{sw}\left(|\theta/\pi-1/2|\right)^\beta$
\citep[][and references therein]{ben05}. The density factors of
$\epsilon_\mathrm{in}$ and $\epsilon_\mathrm{rim}$ produce the hollow structure
of the lobe. In the above parameters, we adopted an $\alpha$ of 1.2, $\beta$ of
1.7, $\gamma$ of 0.7, $\epsilon_\mathrm{in}$ of 0.01, $\epsilon_\mathrm{rim}$ of
0.5, and $R_\mathrm{sw}$ of $7\farcs5$, which fit well the appearance of the
bipolar lobe. The AGB shell exists in the outermost part between the radii of
$R_\mathrm{sw}$ and $R_\mathrm{out}$. The density coefficient $\rho_\mathrm{e}$
is determined with the envelope mass $M_\mathrm{env}$. We applied a gas-to-dust
mass ratio of 160, which is often used for oxygen-rich evolved stars
\citep{knapp85}. The model geometry is viewed at a viewing angle
$\theta_\mathrm{view}$ measured from pole-on. The free parameters and their
ranges are an $R_\mathrm{in}$ of 30, 50, and 100~$R_\mathrm\star$; $H$ of 0.2,
0.3, and 0.4; $M_\mathrm{env}$ of 2, 4, 6, and 8~$M_{\sun}$; and
$\theta_\mathrm{view}$ of 65\degr, 55\degr, and 45\degr.

For the dust grains, we simplify the model as much as possible to focus on
determining the grain sizes which characterize the optical properties of the
dust the most. In our modeling, we assume spherical cores with a mantle and
an MRN-like size distribution of $0.005~\mu$m~$\le a\le a_\mathrm{max}~\mu$m
with $n\left(a\right)\propto a^{-3.5}$ \citep{mrn77}. The $a_\mathrm{max}$
values are different for the disk and the envelope. The chemical compositions
of the core and the mantle are astronomical silicate \citep{draine85} and
crystalline water ice \citep{bertie69}, respectively. The water ice mantle is
assumed to have a constant thickness proportional to the core radius, which is
determined with a volume fraction of the water ice to the grain core
$f_\mathrm{m}$. This value is chosen to be 0.2 which fits the 3~$\mu$m water
ice absorption feature. The grain size in the lobe was estimated from the
polarization values in the lobe, because the NIR polarization strongly depends
on the grain size in cases of submicron size. In our NICMOS2 data, the 2~$\mu$m
polarization is $P=40$ -- 50~\% in the upper lobe and $P=50$ -- 55~\% in the
lower lobe.  With a large inclination angle of $35\degr$, which is expected
from CO observations, the calculated polarization becomes too low in the upper
lobe and too high in the lower lobe.  The average polarization between the
upper and lower lobes for the $a_\mathrm{max}$ of 0.3, 0.5, and 0.7~$\mu$m are
60~\%, 45~\%, and 20~\%, respectively. We apply the $a_\mathrm{max}=0.5~\mu$m
dust model in the envelope in the subsequent models.  The free parameter
is the $a_\mathrm{max}$ of the disk and the values to examine are 0.5~$\mu$m,
1.0~$\mu$m, 10.0~$\mu$m, 100.0~$\mu$m, 1000.0~$\mu$m, and 10\,000.0~$\mu$m.

\subsection{Result}\label{result}
From the aforementioned parameter sets, we selected a good model with the
following parameters: $R_\mathrm{in}=30~R_\star$, $H=0.3$,
$R_\mathrm{disk}=4500$~AU, $M_\mathrm{env}=4~M_{\sun}$,
$R_\mathrm{out}=30\arcsec$, $\theta_\mathrm{view}=55\degr$, and
$a_\mathrm{max}=1000.0~\mu$m.
These are summarized in Table~\ref{i19343_parm}.

The top panels of Fig.\,\ref{rts_i19343} show the cross section of the mass
density distribution and the dust temperature distribution. As seen in the mass
density distribution map, the superwind shell has a waterlily petal-shaped
appearance. In the dust temperature distribution map, intermediate to high
temperature components ($T_\mathrm{d}\ga100$~K) are seen in the innermost
low-density polar region. The rim of the superwind and AGB lobes has
a low temperature of $\la70$~K, which characterizes the flux in the
far-infrared (FIR) or longer wavelengths.

The bottom panels of Fig.\,\ref{rts_i19343} show the opacity (i.e.\,the
size-averaged cross section per particle mass) of the modeled dust and the SEDs.
To show the effect of grain sizes, we also present other dust models with
$a_\mathrm{max}=1.0~\mu$m and $a_\mathrm{max}=10.0~\mu$m for comparison.
In the opacity plot, the difference of the wavelength dependence for different
dust models is clearly visible. As $a_\mathrm{max}$ increases,
(1) the wavelength where the opacity drops becomes longer and (2) the strengths
of water ice and silicate features become weaker. The SED plot compares the
model results of different grain sizes of $a_\mathrm{max}=1.0~\mu$m,
10.0~$\mu$m, and 1000.0~$\mu$m to the observations. The model SEDs reproduce
the water ice features at 3~$\mu$m, 44~$\mu$m, and 62~$\mu$m and the silicate
features at 9.7~$\mu$m and 18~$\mu$m. Because the disk temperature is higher
than the temperature of the superwind and AGB shells, in principle,
the grain-size effect appears in the mid-infrared (MIR) flux, and it is
possible to constrain the grain size by SED fit. On the longer wavelength side,
the difference in the SED appears only for $\lambda\la50~\mu$m.

Figure\,\ref{img_i19343} shows the model results of the intensity images (top
panels) and polarization images (bottom panels). The results of two different
dust models of $a_\mathrm{max}=1.0~\mu$m (middle column) and
$a_\mathrm{max}=1000.0~\mu$m (right column) are compared with the observation
(left column). The intensity images reproduce a waterlily petal-shaped bipolar
appearance reasonably well. In the polarization images, the difference of the
polarization values between the upper and lower lobes is too strong,
as mentioned before. With respect to the effect of the grain size, we do not
find any sufficient difference from our results.

Hence, it is difficult to determine the grain size in the disk of M 1--92 from
the SED, intensity image, and polarization image, which we used in our modeling.
The reason is explained as follows. The MIR flux $F$ from the disk is
approximately proportional to $M_\mathrm{disk}\kappa\left(a\right)$.
Based on our assumption of a constant optical depth of the disk for different
grain models, the optical depth $\tau$ is proportional to
$M_\mathrm{disk}\left(\kappa\left(a\right)+\sigma\left(a\right)\right)$.
Therefore, we obtain $F\propto \tau \kappa\left(a\right)/\left(\kappa\left(a\right)+\sigma\left(a\right)\right)$. For grain models with
$a_\mathrm{max}>10~\mu$m, the $\kappa\left(a\right)/\left(\kappa\left(a\right)+\sigma\left(a\right)\right)$
value becomes nearly constant. With the similar reason, the wavelength
dependence of the dust scattering matrix elements become low in large grains,
resulting that the effect of the grain size does not appear much in polarization
images in the optical to NIR. Thus, the presence of large grains are examined
often using a spectral index of dust opacity $\beta$ in longer wavelengths,
i.e.\,the submillimeter and millimeter wavelength ranges \citep[e.g.][]{jura97}.
In objects with large grains in the disks, the spectral slopes of the flux
in these wavelength ranges are shallower due to lower $\beta$ values
\citep[e.g.][]{draine06}.
However, since $F$ and $\tau$ depend on the disk mass, if the disk mass is
determined by other means, the grain size can be constrained.
\cite{bujarrabal98b} obtained $\sim$$1\arcsec$ resolution images of M 1--92 in
the \element[][13]{CO} $J$=2--1 emission line and an estimate of a gas disk
mass of 0.2~$M_{\sun}$ \citep[see also][]{alcolea07}. The calculated disk
masses in our modeling are 0.022, 0.0098, 0.017, 0.049, 0.15, and
0.47~$M_{\sun}$ for dust models of $a_\mathrm{max}=0.5$, 1.0, 10.0, 100.0,
1000.0, and 10\,000.0~$\mu$m, respectively. We find that the
$a_\mathrm{max}=1000.0~\mu$m model provides the best estimate.
However, facts such as the gas-to-dust mass ratio, a region that is considered
to be the disk, and the dust size distribution can lead some uncertainties in
the total disk mass.  For example, the gas-to-dust mass ratio, which we adopted
in our model is estimated from a number of mass-losing oxygen-rich giant stars
\citep{knapp85}. The value of individual objects varies with a factor of
$\sim$2 in general \citep[see also][]{olofsson93}. In a special case of objects
with a long-lived disk, this value can be $\sim$0.01 times lower \citep{jk99}.
It is not clear whether M 1--92 has such an extra ordinarily low value or not.
However, the calculated disk mass is at least unlikely to be significantly
higher than the aforementioned values.  With models with small grains (micron
size or smaller) in the disk, it is hard to explain the estimated CO disk mass.
Taking into account the possible uncertainty by a factor of $\sim$2 in the
gas-to-dust mass ratio, we conclude that dust in the M 1--92 disk is
significantly larger than in the envelope and the $a_\mathrm{max}$ value is
expected to be larger than $100.0~\mu$m.

\section{Discussion}\label{discuss}
\subsection{Grain size}
For the last 20 years, grain growth in AGB stars and PPNs has been studied and
several observational evidences has been reported. Of these, a particular interest to us lies in a series of studies in carbon stars that show silicate dust
features such as the \object{Red Rectangle}, \object{BM Gem},
\object{V 778 Cyg}, and \object{AC Her}. In this class of objects, the dual
chemistry is explained by the presence of a disk, in which oxygen-rich dust
is stored \citep[][and references therein]{waters98,le90,barnbaum91}. In fact,
some observations have detected disk-like motions in CO emission line
observations \citep[e.g.\,][]{kahane98,fong06} and a spatially resolved
Keplerian rotating motion in the Red Rectangle
\citep{bujarrabal03,bujarrabal05}. If such disks live for a long time, dust
particles can coagulate by grain-grain collision and grow in size
\citep{jura00b,yamamura00}. More recently, aforementioned evidence has been
also found even in some single chemistry systems. For example, \cite{deruyter06}
found flux excesses in $\lambda\ga2.0~\mu$m in their 51 post-AGB stars which
are known to have binary companions. These infrared excesses indicate small
inner radii of dust structures, which are kept closer to the star due to their
rotating motion of the disk rather than their expanding motion. They conclude
that presence of disks are common phenomenon in binary post-AGB stars. Evidence
for large grains has also been found in some PPNs such as \object{AFGL 2688}
\citep{jura00a}, \object{IRAS 17150--3224} \citep{meixner02},
\object{IRAS 18276--1431} \citep{sc07}, \object{IRAS 19475+3119}
\citep{sahai07}, \object{IRAS 22036+5306} \citep{sahai08}, and
\object{IRAS 22272+5435} \citep{ueta01} including our result of M 1--92.
\cite{fong06} performed a survey observation of gas phase kinematics of 38
AGBs to PNs, which include some of aforementioned objects. The objects with
evidence of grain growth, except IRAS 22272+5435 (and IRAS 19475+3119), have
a disk-like motion or a bipolar appearance in the optical and NIR images and
their initial masses are expected to be higher than those of spherical and
elliptic PPNs and PNs with expanding motions (outflow) instead of rotating
motions (disk) \citep{cs95,meixner02}. It is obvious that grain growth is
affected by factors such as disk geometries, kinematics of the disk, evolution
of the stellar system (i.e.\,duration time of the dust processes), and the
stellar masses. Previous work has provided qualitative evaluations to determine
if large grains exist or not and if a disk-like structure exist or not to
explore the possibility of the disk hypothesis and grain growth. Future work
should probably be focused on more qualitative, direct discussions on the disk
hypothesis. For this purpose, a more systematic and self consistent analysis
are required to estimate the physical parameters of the disk and dust.

\begin{figure*}
  \sidecaption
  \includegraphics[width=12cm]{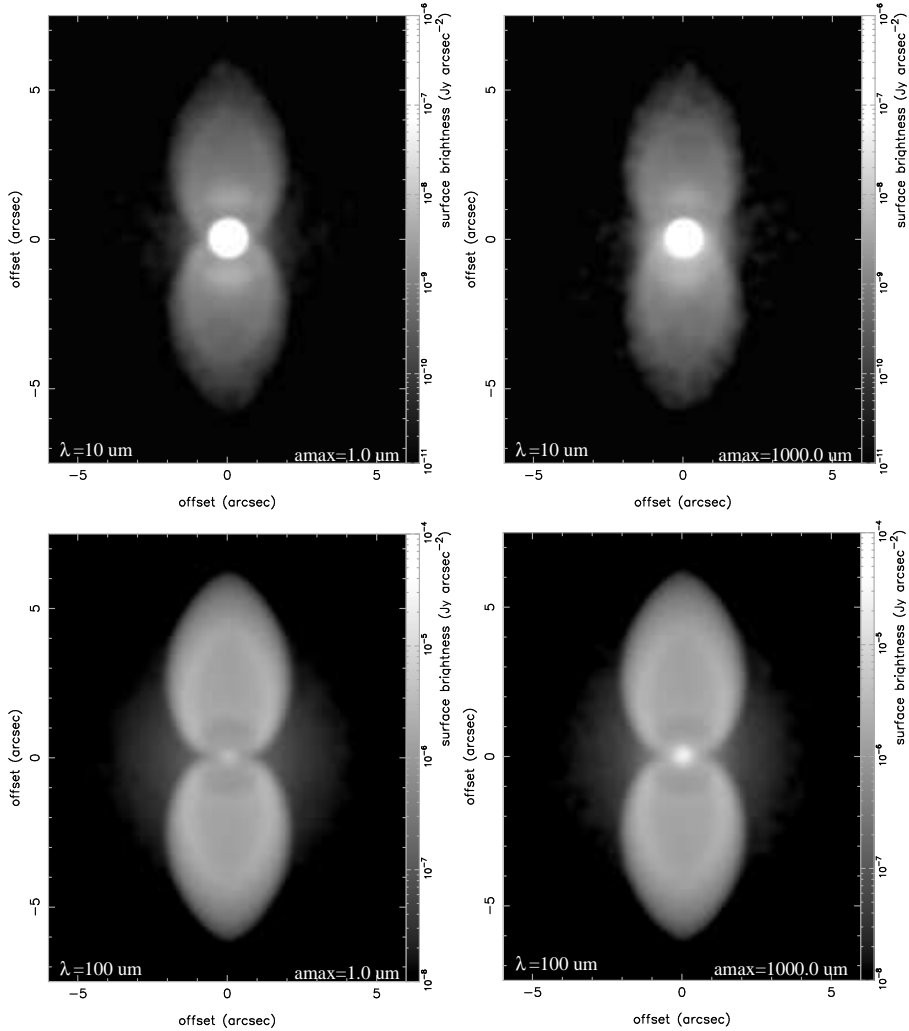}
  \caption{Model results of thermal emission images. The top and bottom panels
           show images at the wavelength of 10.0~$\mu$m and 100.0~$\mu$m, and
           left and right columns correspond to dust models of
           $a_\mathrm{max}=1.0~\mu$m and $a_\mathrm{max}=1000.0~\mu$m in the
           disk, respectively. Each image is convolved with a Gaussian function
           with a 0\farcs3 FWHM.
         }
  \label{ems_i19343}
\end{figure*}

With respect to dust, the presence of large dust grains in the nebulae is also
intriguing. Optical to NIR imaging polarimetry offers a great advantage in this
issue. In polarimetric images of circumstellar dust shells around AGB and
post-AGB stars, a centro-symmetrically aligned polarization vectors are
seen. If the dust shell geometry and the chemical composition of the dust are
known, the degree of polarization would provide a good estimate of the grain
size. Particularly, in an optically thin regime where most light from the
central star is scattered only once by dust in the nebula, the upper limit can
be provided \citep[e.g.][]{dougados90,pendleton90}. It is possible to obtain
a rough estimate even from a single wavelength data, under some circumstances,
which is an advantage of polarimetric images compared to intensity images,
visibility data, and SEDs. In most cases including our sample, intermediate to
high polarizations ($P\ge20$~\%) in the NIR are detected and indicate submicron
sizes in the shell. The dust in nebulae in large distances from the central
star is thought to be ejected in the AGB mass-loss wind with mainly expanding
motions instead of orbiting motions around the central stars. Thus, these grains
are not expected to undergo physical processes such as grain-grain collision
and ice formation on the grain core much. Thus, dust in this region is likely to
preserve the original size distribution, which is determined in the dust
formation process, as explained by the dust formation and stellar wind theories
\citep[e.g.][]{gl04,hoefner08}. In previous observations, we did not find any
significant difference in dust sizes or at least in the polarization values in
various object classes \citep[e.g.][UMM07]{jura96}. It appears that the initial
mass of the stars (or the luminosity) and the chemical composition of the
stellar system (i.e.\,oxygen-rich or carbon-rich) do not affect the dust
particle size in the dust formation much.

\subsection{Thermal emission images}
In this section, we present some model images in the MIR and FIR.
As is expressed in the radiative transfer theory, flux from objects in these
wavelength regions (thermal emission regime) is characterized by a combination
of the dust temperature and density distributions and the dust absorptive
opacity, while optical and NIR fluxes are governed by scattered light.
It is worth to take multiple wavelength data into account, which is based on
different physics, to better constrain the model parameters. Unfortunately,
observed images of M 1--92 in the MIR and FIR are not publicly available at
this time. Thus, we use the model results to predict what we see in the
nebulosity and what parameters can be constrained.

Figure\,\ref{ems_i19343} shows $\lambda=10.0~\mu$m and 100.0~$\mu$m images
(top and bottom panels, respectively) of our selected model results with
different grain populations of $a_\mathrm{max}=1.0~\mu$m and 1000.0~$\mu$m in
the disk (left and right panels, respectively). In the 10.0~$\mu$m images, the
surface brightness of the nebulosity is fainter by a factor of $<$10$^{-5}$
than the peak surface brightness of the central source. On the other hand,
these are comparable in the 100.0~$\mu$m images. We find a clear correlation
between the model images and the dust temperature and density distribution maps
(Fig.\,\ref{rts_i19343}). The appearance of the bright central source in the
10.0~$\mu$m images is due to the presence of a warm (T$\ga$300~K), high-density
dust region (i.e.\,inner part of the disk). In fact, such results were also
reproduced by previously reported modeling
\citep[e.g.][]{lopez00,meixner02,um03}.
As explained in detail in these previous papers, in the bipolar case, i.e.
optically thick case at a given wavelength, the inner part and polar region of
the dust shell are heated well. Taking into account the dust density, MIR
photons are mainly emitted from a central, compact region, but only a little
from an envelope far from the central star. We illustrate the effect of dust
temperature in the 100.0~$\mu$m images by comparing with the 10.0~$\mu$m images.
As is expected in the radiative transfer theory, FIR photons are emitted more
efficiently from cold dust (T$\sim$30~K) than warm dust. In the other words,
a longer wavelength image traces generally more of the outer region. Therefore,
in spite of a high density in the disk, the central source is faint in this
wavelength. In contrast, the superwind shell, where cold dust exists, becomes
brighter. In terms of the total flux, i.e.\,the flux seen in the SED plot,
the total mass of the geometry components are also important because the flux
is proportional to the dust mass. The estimated mass of the superwind shell is
2.6~$M_{\sun}$, which is much higher than the disk mass of 0.15~$M_{\sun}$.
The FIR flux is governed much more by the flux from the superwind than the disk.
The FIR data is useful for additional constraining of the dust mass of the
superwind and AGB wind shell; however, it is unrealistic to probe the detailed
structure using the present imagers because of a limitation of angular
resolutions (a few arcsec).

The surface brightness of the central source would be useful to constrain the
grain sizes in the disk. The 10.0~$\mu$m images show a brighter central source
in the $a_\mathrm{max}=1000.0~\mu$m dust model than in the
$a_\mathrm{max}=1.0~\mu$m dust model. As seen in Fig.\,\ref{rts_i19343},
the 10~$\mu$m silicate feature (or opacity) becomes weaker in larger grains
than in smaller grains. This results in lower extinction in larger grains.
However, the wavelength dependence of the dust opacity becomes similar for
large grains. In reality, we do not expect sufficient accuracy to determine
the grain size with observed MIR$/$FIR images, if the $a_\mathrm{max}$ value
exceeds $\sim$10.0~$\mu$m. Another possible method, besides the one we proposed
in this paper, is to use the flux excess in the submillimeter and millimeter
wavelengths. In these wavelength ranges, the flux from the superwind shell
decreases by an order compared to FIR flux, as is expected in the SED, but flux
from the disk does less if the dust in the disk is sufficiently grown in size
(the spectral opacity index $\beta<1$). In further works, one can perform high
angular resolution imaging in these wavelegnth ranges to better constrain the
grain sizes and disk masses.

\section{Conclusion}
We performed two-dimensional radiative transfer modeling of the dust shells of
the bipolar PPN M 1--92. Our modeling applied geometries with a disk and
bipolar envelope surrounded by an AGB wind shell, each of which has different
dust characteristics. The model parameters were constrained by comparing them
with the previously observed SEDs, the intensity and polarization images from
the HST NICMOS 2 archived data (UMM07), and a previous radio observation in the
CO emission line \citep{bujarrabal98b,alcolea07}. With a waterlily-shaped
hollow envelope, the bright bipolar lobes of M 1--92 were reproduced. For the
dust sizes, we found submicron-sized grains ($a_\mathrm{max}=0.5~\mu$m) in the
bipolar lobes.  The dust size in the disk was constrained with the disk mass,
which was estimated from the CO emission line data. We obtained
$a_\mathrm{max}=1000.0~\mu$m. Although this estimate includes some uncertainties
in the gas-to-dust mass ratio, the grain size distribution, and the geometry of
the region considered to be the disk, an $a_\mathrm{max}\la100.0~\mu$m is hard
to explain the disk mass and a significantly large size of $>100.0~\mu$m is
expected.  We conclude that grain growth is likely to occur in the M 1--92 disk.
Further works including submillimeter and millimeter wavelength imaging and
photometry would provide better interpretations on grain growth.  The small
grains in the bipolar lobes are consistent with results of many other AGB and
PPNs, suggesting that dust formed by AGB mass loss has a similar size, which
does not depend on the object much. The formation of such submicron-sized dust
is explained with dust formation and stellar wind theories.  The presence of
large grains in the disk should be explained with another mechanism, which is
probably the long-lived disk hypothesis. The grain growth possibly depends on
the disk geometry and the stellar temperature. Better understanding will be
provided if a detailed systematic analysis is made for many other samples.


\end{document}